\documentclass[11pt,twoside]{article}

\usepackage{asp2006}
\usepackage{epsf}
\usepackage{lscape}

\begin{document}
\title{Asymmetric Warfare: M31 and its Satellites}   %%% Fill in title
\author{Mark A. Fardal}   %%% Fill in author names
\affil{Dept. of Astronomy, University of Massachusetts, Amherst, MA 01003}    %%% Fill in author affiliations

\begin{abstract} %%% Abstract to run on from here.
Photometric surveys of M31's halo vividly illustrate the wreckage
caused by hierarchical galaxy formation.  Several of M31's satellites
are being disrupted by M31's tidal field, among them M33 and And I,
while other tidal structures are the corpses of satellites already
destroyed.  The extent to which M31's satellites have left battle
scars upon it is unknown; to answer this we need accurate orbits and
masses of the perturbers.  I focus here on M31's 150-kpc-long Giant
Southern Stream (GSS) as an example of how these can be determined
even in the absence of a visible progenitor.  Comparing N-body models
to photometric and spectroscopic data, I find this stream resulted
from the disruption of a large satellite galaxy by a close passage about 
750 Myr ago.  The GSS is connected to several other debris structures
in M31's halo.  Bayesian sampling of the simulations estimates the
progenitor's initial mass as $M_\star = 10^{9.5 \pm 0.2} M_{\sun}$, showing it
was one of the most massive Local Group galaxies until quite recently.
The stream model constrains M31's halo mass to be $(1.8 \pm 0.5)
\times 10^{12} M_{\sun}$.  While these small uncertainties neglect several
important degrees of freedom, they are likely to remain good even
with a more complete model.  Future work on M31's satellites and
streams will provide independent constraints on M31's 
mass, and reveal the shared history of M31 and its halo components.
\end{abstract}
\section{Introduction}
Despite the term ``Local Group'',
M31 has never interacted with the Milky Way and the two galaxy virial
radii are far from touching.  So M31 is really just an 
isolated spiral galaxy that happens to be close enough to study
in great detail.  Of course
even ``isolated'' galaxies have their satellites; M31 has two nearby ones
at around 1\% of its own stellar mass (M32 and NGC 205), while the spiral M33
has around 10\% of M31's mass but at such large distance that it is
unclear whether it is truly a bound satellite.  M31's disk is
far from regular: it is significantly warped, and star formation 
indicators show a fairly well-defined 10 kpc ring that 
dominates the total star formation rate.
Various authors have attributed features within M31 to the impact of
satellites including M32 and NGC 205 \citep{gordon06}.
In fact, \citet{block06} assigned the entire 10 kpc ring to a
Cartwheel-like expanding wave, requiring a collision 250 Myr ago which
they attribute to M32.  However, none of the satellite orbits are
currently known, so all of these scenarios are highly speculative.

\begin{figure}[!ht]
%\plotone{image_coarse_labeled_h.eps}   %lowres for web distribution
\plotone{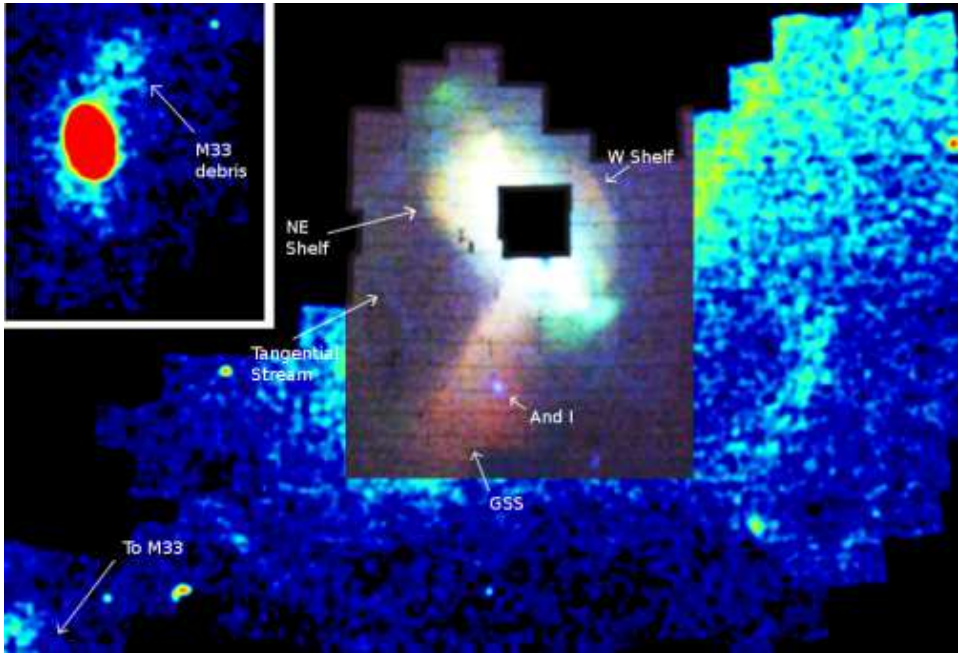}   %lowres for web distribution
%\plotone{image_.eps}    %highres for print
\caption{
\label{fig:maps} 
A $20 \times 14$ degree view of M31
($272 \times 191$ kpc assuming a distance of 780 kpc)
from the PAndAS survey \citep[][maps by A. McConnachie 
and N. Martin]{mcconnachie09}.  
The inner map shows RGB branch color 
while the outer map shows false-color density.
An inset shows M33 on the same scale.
The image is made from matched-filter maps on a tangent plane
projection 
which are sensitive to RGB stars at the distance of M31.
Each individual field is 1 degree across.
Visible rectangular artifacts are consequences of chip gaps and 
variations in field quality.
The GSS and the NE and W shelves
are all metal-rich features as indicated by the red color of their 
red giant branch.  Metal-poor features include several dSph galaxies,
the tangential arc to the E connecting to the dSph And I 
(dot in the middle of the GSS),
and a radial stream along the NW edge of the survey at the upper right.
}
\end{figure}

Photometric surveys of M31's halo have discovered a great
deal of stellar debris, some likely from M31's disk and some tracing
past or ongoing mergers.  The recent PAndAS survey
\citep[][Figure 1]{mcconnachie09} extends to M33 and 
will eventually fill in a 150 kpc circle around M31, 
down to stellar magnitudes of $i_{AB} = 24.5$ at $S/N = 10$.  
Visible in Figure 1 are the Giant Southern Stream
\citep{ibata01}, as well as newly discovered tangential
and radial streams to the E and NW respectively. 
Ledges or ``shelves'' on the NE and W sides are also apparent.  
These faint features, only made apparent by counting individual stars,
are beginning to illuminate the orbits of objects in M31's halo.

\section{The Giant Southern Stream}
\subsection{Observations}
I now focus on the GSS as the best-studied of
the tidal features.  Metallicity estimates of [Fe/H] $\sim -0.5$
suggest the stream originates from a disrupted dwarf galaxy of
stellar mass $M_{sat} \sim 10^9 M_{\sun}$ \citep{font06}.  
From HST photometry that reaches the main sequence, \citet{brown06a,brown06b}
infer a mean age of 8 Gyr and a total lack of stars younger than 4 Gyr.
Distances from the RGB tip magnitude show
the stream's southern end lies far beyond M31 \citep{mcconnachie03}.
Spectroscopic surveys measure the rate at which stream stars speed up
as they fall back into M31's \citep{ibata04,raja06,kalirai06}.
The increasing surface brightness as the stream approaches M31 center
strongly suggests the main body of the stream's progenitor lies
further ahead in the orbit, but where is it?  Early suggestions
included M32 \citep[][but the stream velocities rule out a
simple connection]{ibata01}, the NE Shelf \citep{ferguson02}, or a
stream evident in planetary nebulae \citep{merrett03}.

\begin{figure}[!ht]
\plottwo{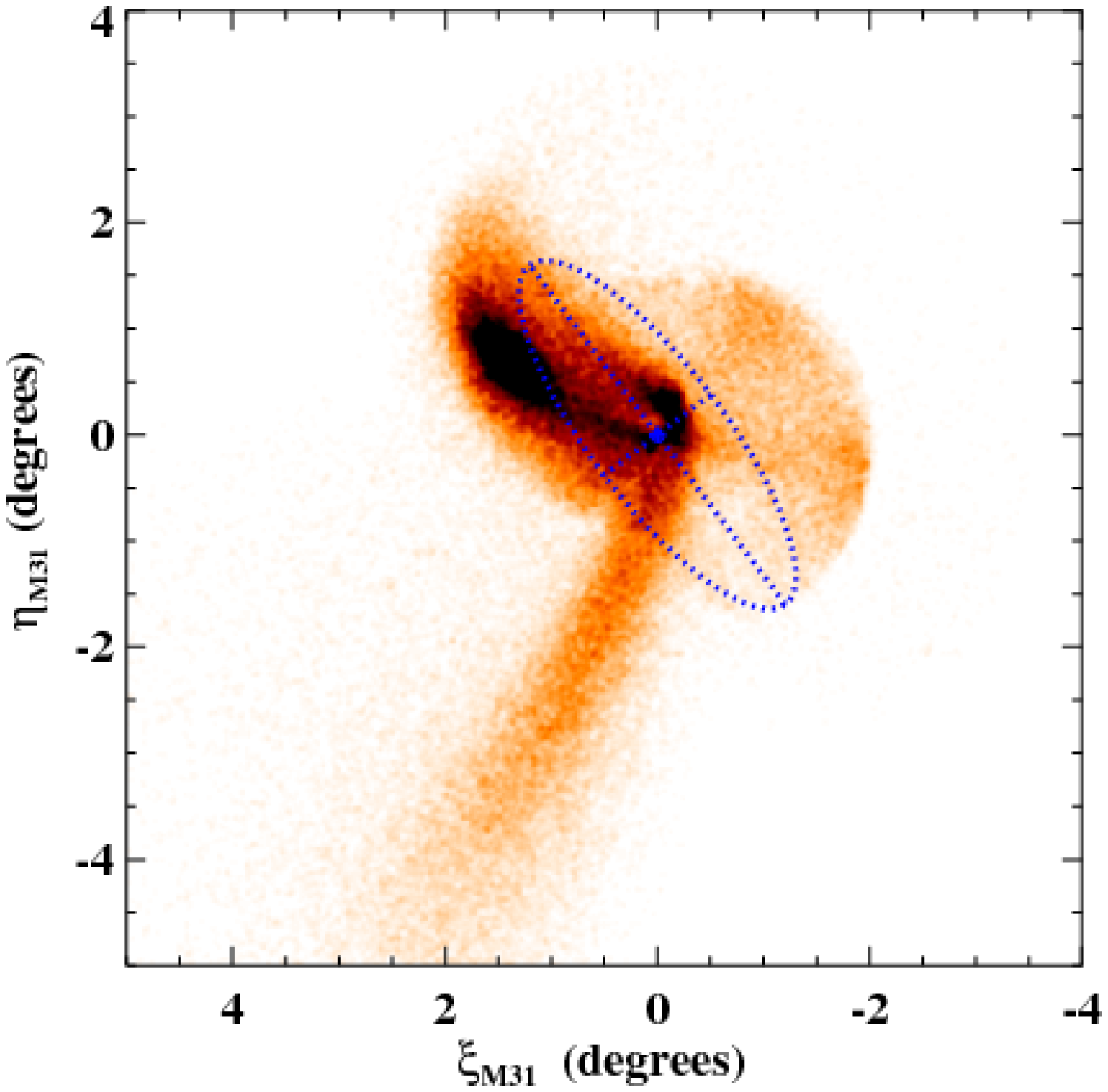}{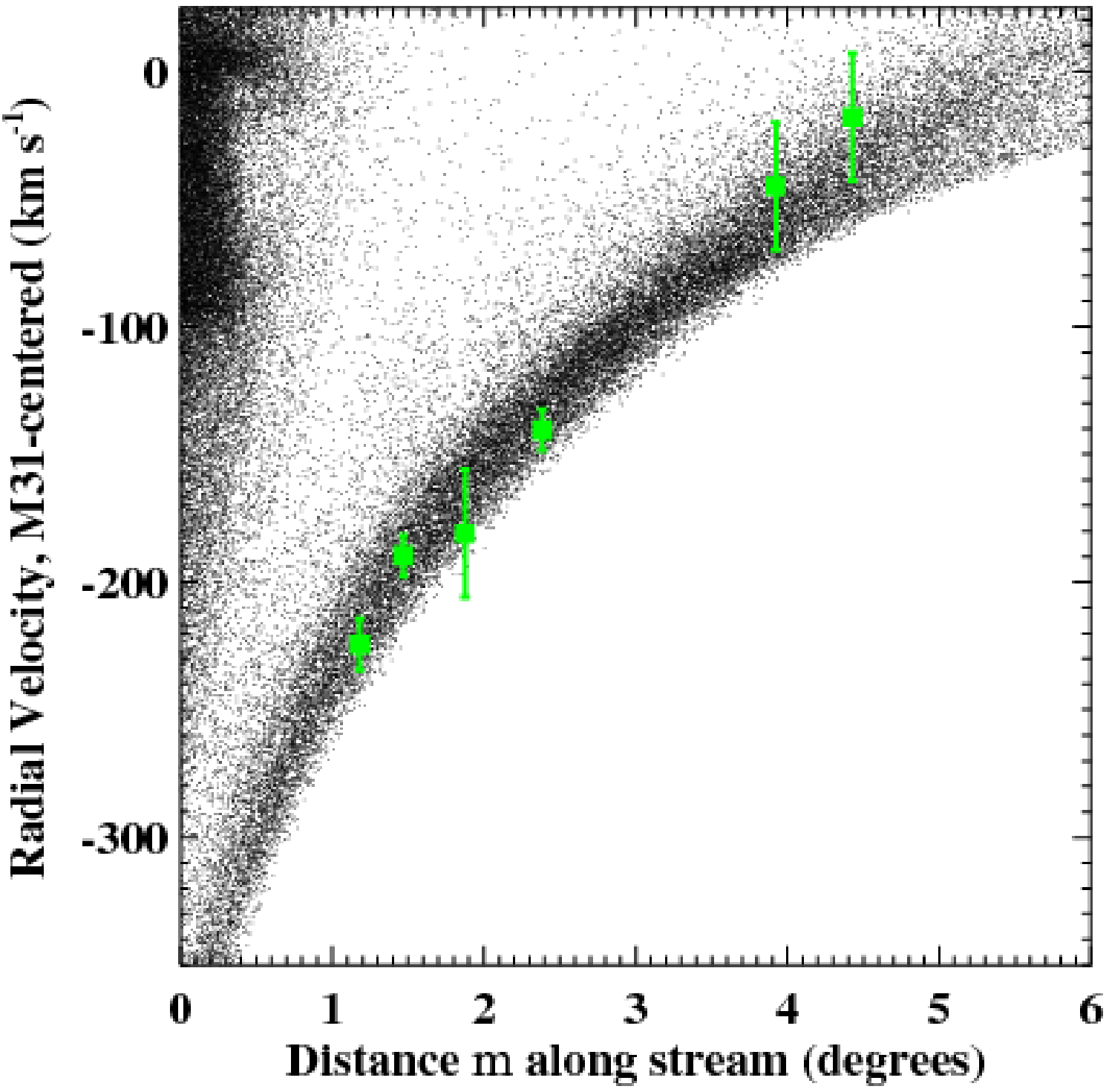}   %lowres for web distribution
%\plottwo{shelfim.eps}{streamvel.eps}    %vector for print
\caption{
\label{fig:shelf} 
Left: Surface density of N-body particles 
in a tangent plane projection, in one sample from parameter space
(resimulated with more particles for visual clarity).
Note the similarity to Figure 1's GSS and shelves.
Right: the radial velocity of N-body particles versus distance along
the stream, from the same model.  The GSS from the left 
panel appears as a cold diagonal component with velocities similar
to the observed mean values, from \citet{ibata01,raja06,kalirai06}.
}
\end{figure}

\subsection{Modeling}
The earliest modeling work on the GSS concentrated on fitting orbits
through the stream.  However, our initial work combining orbits with
N-body simulations demonstrated that {\em the stream does not follow
the orbit}. The stream consists of stars lifted to much higher orbital
energies than that of the GSS progenitor, increasingly so as the
radius increases, and the progenitor typically will have
apocenter at half of the stream's current length.  In our first
modeling attempt \citep{fardal06} we provided an approximation connecting
the stream and the orbit, and used this to show the stream must
continue somewhere to the NE of M31's center.  Later \citep{fardal07} we
provided a scenario in which the subsequent loops of the stream create
both the NE and W shelves (Figure 2).  This model also explains the
velocity trend and spatial location of the PNe ``stream'', and shows
this ``stream'' is simply a subset of the NE shelf debris with
coherent velocities resulting from a caustic feature in observed phase
space.  This model has since been strengthened by the strikingly
similar color-magnitude diagrams within the stream and shelf regions
\citep{richardson08}, and the apparent detection of the stream's 
fourth orbital wrap \citep{gilbert07}.
The model can be refined by adding rotation to the progenitor,
which improves agreement with the 
transverse profile of the stream \citep{fardal08}.

Our modeling work takes the progenitor to be mostly stellar and starts
the collision half an orbit before disruption, neglecting the
question of how the satellite gets to that state.  In tests
with live halos, it seems plausible for the satellite to be dragged in
by means of dynamical friction as it loses its halo to tidal stripping.  
Alternatively, an interaction with a perturber such as M33 may be 
responsible for putting it on a highly destructive orbit.

\subsection{Parameter Sampling}
Our model of the GSS appears to meet multiple tests; can we be
more quantitative about its implications?
We have begun a program of Bayesian sampling of parameter space,
within the scenario just outlined.  We use around 30,000
N-body simulations to sample a likelihood function formed from
the stream's position, distance, and velocity, plus the 
brightness of several bins within the GSS proper and the W shelf.
The brightness ratio of the GSS and W shelf turns out to be 
quite sensitive to orbital phase, which puts strong constraints 
on the model.
We run Markov chains in parallel until they converge to equilibrium.
The satellites are simple Plummer-model progenitors disrupting within the
best-fit M31 potential family of \citet{geehan06}.  That model contains 
large, quantified uncertainty in the halo mass; the new constraints
from the GSS reduce this uncertainty significantly.

At this stage we use a restricted parameter space, 
eliminating five of the six orbital parameters by fixing them to their
best-fit values as a function of orbital phase.  We then sample from
the dimensions of orbital phase, halo mass, and progenitor
satellite mass.  These parameters are very well constrained, 
with orbital phase $t/P = 1.25 \pm 0.15$ periods past disruption
at pericenter $760 \pm 20$~Myr ago,
virial mass $M_{100} = 10^{12.26 \pm 0.12} M_{\sun}$ 
(within $R_{100}$ which encloses 100 times the closure density),
and GSS progenitor mass $M_{sat} = 10^{9.56 \pm 0.22} M_{\sun}$.
Figure 3 shows these distributions.  The eliminated orbital degrees of
freedom were already well constrained at a given orbital phase, so we
do not expect their restoration to widen the error bars dramatically.
We have not yet estimated systematic errors, from M31's distance,
radial halo profile, or halo shape for example, but even so
it is clear the GSS puts useful constraints on M31's properties.

\begin{figure}[!ht]
\plottwo{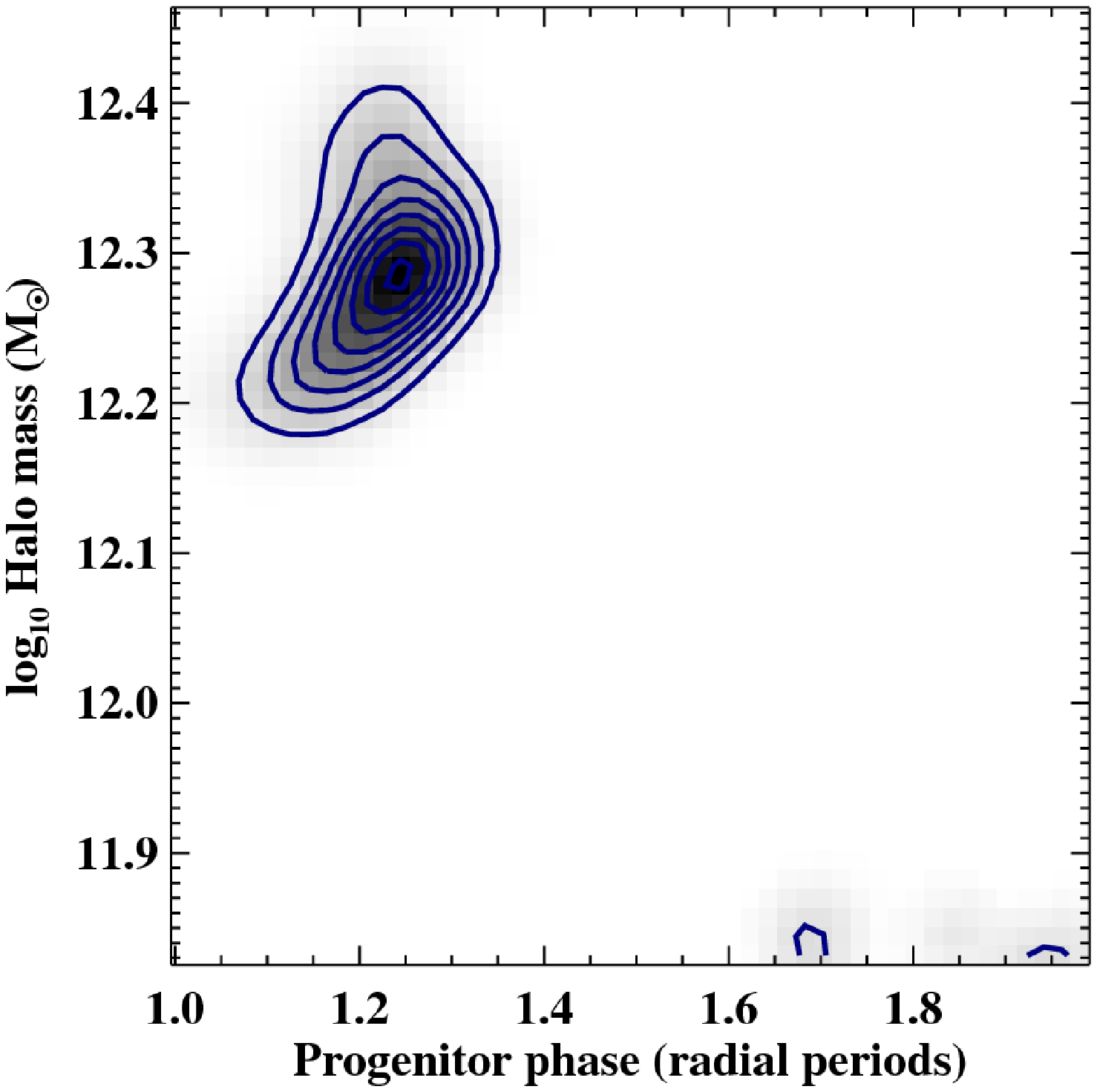}{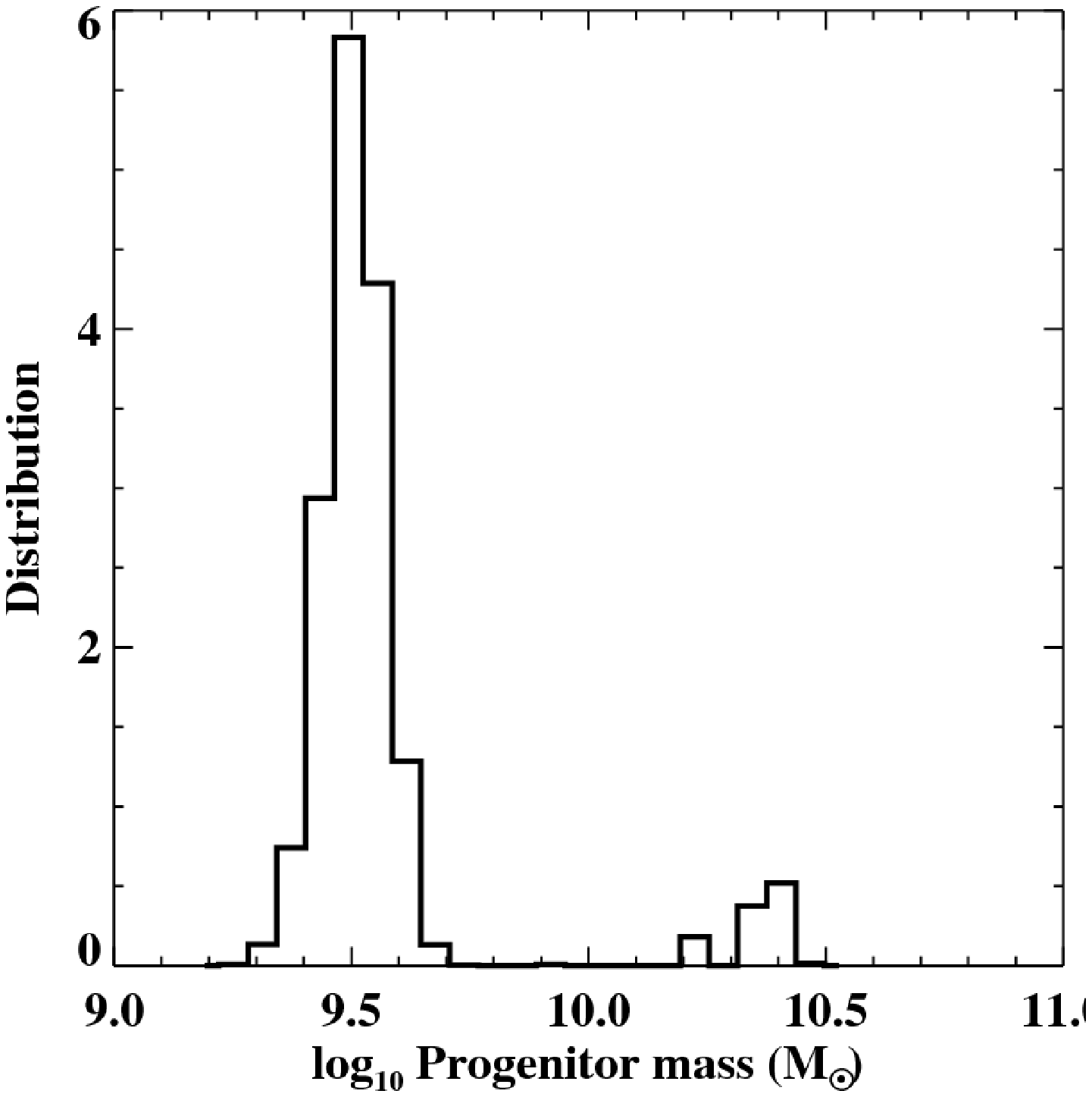}  %lowres for web distribn
%\plottwo{phase_hmass.eps}{stellarhist.eps}           %vector for print
\caption{
\label{fig:histograms} 
Left: contours of halo mass $M_{100}$ versus 
the progenitor's current orbital phase, from our Bayesian sampling.
Right: histogram of the progenitor mass, which is taken to be purely stellar.
}
\end{figure}

\subsection{Implications}
This modeling work has a number of implications:
\begin{itemize}
\item The GSS is due to a previously unknown satellite of M31, not
any of the currently intact ones such as M32.  
\item Its disruption took place 760~Myr ago, small in
cosmological terms.  This is long after the last recorded
star formation in the progenitor, but is far too early to induce the
expanding star-forming wave envisioned by \citet{block06}. (Indeed it 
is unclear whether the ring is expanding, it may be a static feature).
\item The GSS progenitor's stellar mass of $10^{9.5} M_{\sun}$,
which is in good agreement with the measured stream metallicity,
puts it just behind the LMC in the catalogue of Local Group galaxies.
Limits on the heating of M31's disk \citep{mori08} imply any 
dark halo in the progenitor was quite minor at the time of disruption.
\item We measure M31's halo mass at 
$M(r_{100}) = (1.8 \pm 0.5) \times 10^{12} M_{\sun}$. 
This mass implies that M33 is very likely to be a bound satellite.
Our estimated mass is larger than the most likely values 
7--$10 \times 10^{11} M_{\sun}$ from kinematics of
M31's satellites \citep{evans00}, but smaller than the 
value $2.8 \times 10^{12} M_{\sun}$ inferred from the
timing argument \citep{li08}, and lower than expected from 
comparisons of the (observed) stellar and (theoretical) halo 
cosmic mass functions \citep{yang08}.
\item More generally, N-body simulations can be used effectively to
sample the entire parameter distribution in an automated way, at least
if the problem has a single well-constrained mode as seems the case here.  
This shows one can
obtain well-specified parameter estimates, errors, and covariances, even
in cases where simulations are necessary to estimate the observables.

\end{itemize}

\section{Other Satellites}
M33's tidal disturbances in HI have been long known, but similar
disturbances in the stars are detected for the first time in PAndAS
(Figure 1).  Their presence suggests the material is not falling in
for the first time but results from a prior interaction with M31
\citep[][Dubinski et al in prep.]{mcconnachie09}.  Our recent paper
outlines a scenario suitable for exciting disturbances of the observed
size scale in M33.  M33 is placed on an orbit with orbital period 
1.7~Gyr.  After one pass through pericenter at 53 kpc, M33 then
passes through apocenter at 264 kpc and arrives at its present-day
position with tidal distortions comparable in size to those seen in 
the data.  Interestingly, despite the large pericenter, M31's disk shows
some warping and other disturbances from M33's passage.

A tidal stream that seems to emanate from And I is also prominent in
Figure 1.  Our eventual goal is to generate reliable orbital scenarios
for each of the tidal structures visible in Figure 1, using a
combination of photometric and kinematic data.  Use of a Bayesian
formalism as in Section 2.3 will then result in well-understood
uncertainties and multiple independent determinations of M31's halo
mass.  As the understanding of the star formation pattern and history
within M31's disk improves, we will be able to correlate this with the
orbits and determine the extent to which M31's satellites have left
marks upon it.

\acknowledgements I thank my collaborators on the work described here,
including Arif Babul, Raja Guhathakurta, Alan McConnachie,
Mike Irwin, Martin Weinberg, John Dubinski, Larry Widrow, 
and the other members of the SPLASH and PAndAS collaborations.

%%% THE BIBLIOGRAPHY
%%%
%%% CONSULT SECTION 3 OF "INSTRUCTIONS FOR AUTHORS" FOR HOW TO USE NATBIB.
%%% AUTHORS ARE ENCOURAGED TO USE EITHER THE "THEBIBLIOGRAPY" ENVIRONMENT
%%% BY UNCOMMENTING (DELETING THE "%" SYMBOL) THE COMMANDS BELOW, OR BY
%%% USING THE BIBTEX ENVIRONMENT. TO FIND OUT WHICH IS APPLICABLE TO YOUR
%%% CONTRIBUTION, CONSULT THE VOLUME EDITORS FOR YOUR PROCEEDINGS.
%%%

\end{document}